\documentclass[amsmath,aps,prd,showpacs,a4paper,10pt]{revtex4}

 \usepackage{epsf}
 \usepackage{graphicx}    

\begin{document}

 \newcommand{\be}[1]{\begin{equation}\label{#1}}
 \newcommand{\ee}{\end{equation}}
 \newcommand{\bea}{\begin{eqnarray}}
 \newcommand{\eea}{\end{eqnarray}}
 \def\disp{\displaystyle}

 \def\gsim{ \lower .75ex \hbox{$\sim$} \llap{\raise .27ex \hbox{$>$}} }
 \def\lsim{ \lower .75ex \hbox{$\sim$} \llap{\raise .27ex \hbox{$<$}} }

 \begin{titlepage}

 \begin{flushright}
 arXiv:0707.4052
 \end{flushright}

 \title{\Large \bf Interacting Agegraphic Dark Energy}

 \author{Hao~Wei\,}
 \email[\,email address:\ ]{haowei@bit.edu.cn}
 \affiliation{Department of Physics, Beijing Institute
 of Technology, Beijing 100081, China\\
 Department of Physics and Tsinghua Center for
 Astrophysics,\\ Tsinghua University, Beijing 100084, China}

 \author{Rong-Gen~Cai\,}
 \email[\,email address:\ ]{cairg@itp.ac.cn}
 \affiliation{Institute of Theoretical Physics, Chinese Academy
 of Sciences, P.O. Box 2735, Beijing 100190, China}

 \begin{abstract}\vspace{5mm}
 \centerline{\bf ABSTRACT}\vspace{2mm}
A new dark energy model, named ``agegraphic dark energy'',
 has been proposed recently, based on the so-called
 K\'{a}rolyh\'{a}zy uncertainty relation, which arises from
 quantum mechanics together with general relativity. In this
 note, we extend the original agegraphic dark energy model by
 including the interaction between agegraphic dark energy
 and pressureless (dark) matter. In the interacting agegraphic
 dark energy model, there are many interesting features
 different from the original agegraphic dark energy model
 and holographic dark energy model. The similarity and
 difference between agegraphic dark energy and holographic
 dark energy are also discussed.
 \end{abstract}

 \pacs{95.36.+x, 98.80.Qc, 98.80.-k}

 \maketitle

 \end{titlepage}

 \renewcommand{\baselinestretch}{1.6}



\section{Introduction}\label{sec1}

Dark energy~\cite{r1} has been one of the most active fields in
 modern cosmology since the discovery of accelerated expansion
 of our universe~\cite{r2,r3,r4,r5,r6,r7,r8,r9}. The simplest
 candidate of dark energy is a tiny positive cosmological
 constant. However, as is well known, it is plagued by the
 so-called ``cosmological constant problem'' and ``coincidence
 problem''~\cite{r1}. The cosmological constant problem is
 essentially a problem in quantum gravity, since the
 cosmological constant is commonly considered as the vacuum
 expectation value of some quantum fields. Before a completely
 successful quantum theory of gravity is available, it is more
 realistic to combine quantum mechanics with general relativity
 directly.

Following the line of quantum fluctuations of spacetime,
 in Refs.~\cite{r12,r11,r17},
 by using the so-called K\'{a}rolyh\'{a}zy relation
 $\delta t=\lambda t_p^{2/3}t^{1/3}$~\cite{r10} and
 the well-known time-energy uncertainty relation
 $E_{\delta t^3}\sim t^{-1}$, it was argued
 that the energy density of metric fluctuations of
 Minkowski spacetime is given by
 \be{eq3}
 \rho_q\sim\frac{E_{\delta t^3}}{\delta t^3}\sim
 \frac{1}{t_p^2 t^2}\sim\frac{m_p^2}{t^2}.
 \ee
 We use the units $\hbar=c=k_B=1$ throughout, whereas
 $l_p=t_p=1/m_p$ with $l_p$, $t_p$ and $m_p$ being the reduced
 Planck length, time and mass, respectively. It is worth
 noting that in fact the K\'{a}rolyh\'{a}zy relation and the
 corresponding energy density~(\ref{eq3}) have been
 independently rediscovered later for many times in the
 literature (see e.g.~\cite{r30,r31,r32}).

In~\cite{r17}, one of us~(R.G.C.) proposed a new dark energy model
 based on the energy density~(\ref{eq3}). As the most natural
 choice, the time scale $t$ in Eq.~(\ref{eq3}) is chosen to
 be the age of our universe
 \be{eq4}
 T=\int_0^a\frac{da}{Ha},
 \ee
 where $a$ is the scale factor of our universe; $H\equiv\dot{a}/a$
 is the Hubble parameter; a dot denotes the derivative with respect
 to cosmic time. Therefore, we call it ``agegraphic'' dark
 energy. The energy density of agegraphic dark energy is
 given by~\cite{r17}
 \be{eq5}
 \rho_q=\frac{3n^2m_p^2}{T^2},
 \ee
 where the numerical factor $3n^2$ is introduced to parameterize
 some uncertainties, such as the species of quantum fields in
 the universe, the effect of curved spacetime (since the energy
 density is derived for Minkowski spacetime), the non-saturation
 of the quantum fluctuations, and so on. Obviously,
 since the present age of the universe $T_0\sim H_0^{-1}$ (the
 subscript ``0'' indicates the present value of the corresponding
 quantity; we set $a_0=1$), the present energy density of the
 agegraphic dark energy explicitly meets the observed value,
 provided that the numerical factor $n$ is of order
 unity. In addition, by choosing the age of the universe rather
 than the future event horizon as the length measure, the drawback
 concerning causality in the holographic dark energy model
 does not exist in the agegraphic dark energy model~\cite{r17}.

Considering a flat Friedmann-Robertson-Walker~(FRW) universe with
 agegraphic dark energy and pressureless matter, the corresponding
 Friedmann equation reads
 \be{eq6}
 H^2=\frac{1}{3m_p^2}\left(\rho_m+\rho_q\right).
 \ee
 It is convenient to introduce the fractional energy densities
 $\Omega_i\equiv\rho_i/(3m_p^2H^2)$ for $i=m$ and $q$. From
 Eq.~(\ref{eq5}), it is easy to find that
 \be{eq7}
 \Omega_q=\frac{n^2}{H^2T^2},
 \ee
 whereas $\Omega_m=1-\Omega_q$ from Eq.~(\ref{eq6}). By using
 Eqs.~(\ref{eq5}), (\ref{eq7}) and the energy conservation
 equation $\dot{\rho}_m+3H\rho_m=0$, we obtain the equation
 of motion for $\Omega_q$~\cite{r17},
 \be{eq8}
 \Omega_q^\prime=\Omega_q\left(1-\Omega_q\right)
 \left(3-\frac{2}{n}\sqrt{\Omega_q}\right),
 \ee
 where a prime denotes the derivative with respect to the
 $e$-folding time $N\equiv\ln a$. From the energy conservation
 equation $\dot{\rho}_q+3H(\rho_q+p_q)=0$, as well as
 Eqs.~(\ref{eq5}) and~(\ref{eq7}), it is easy to find that the
 equation-of-state parameter~(EoS) of the agegraphic dark
 energy, $w_q\equiv p_q/\rho_q$, is given by~\cite{r17}
 \be{eq9}
 w_q=-1+\frac{2}{3n}\sqrt{\Omega_q}.
 \ee
 Obviously, the EoS of the agegraphic dark energy is always
 larger than $-1$, and cannot cross the so-called phantom divide
 $w_{de}=-1$. The total EoS
 $w_{tot}\equiv p_{tot}/\rho_{tot}=\Omega_q w_q$. To accelerate
 the expansion of our universe, $w_{tot}<-1/3$ is necessary.
 Thus, $n>2\Omega_q^{3/2}(3\Omega_q-1)^{-1}$ follows. It is
 easy to see that if $\Omega_q>1/3$~(nb. $\Omega_q\simeq 0.7$
 today), the minimum of $2\Omega_q^{3/2}(3\Omega_q-1)^{-1}$
 is $1$ at $\Omega_q=1$. Therefore, $n>1$ is necessary to drive
 the (present) accelerated expansion of our universe.

In addition, it is of interest to compare Eqs.~(\ref{eq8})
 and~(\ref{eq9}) with the ones of the holographic dark
 energy~\cite{r13,r15,r18}. Obviously, they are fairly similar.
 Of course, there are some differences. Except for the slight
 differences of the numerical constant, the most important
 difference is the sign before the term $\sqrt{\Omega_{de}}$~(the
 subscript $de=q$ and $\Lambda$ for the agegraphic dark energy
 and holographic dark energy, respectively). In fact, this sign
 is opposite in these two models. We will see that this
 difference brings about some interesting features to the
 agegraphic dark energy.

In fact, soon after the appearance of~\cite{r17}, it is found
 that the agegraphic dark energy model cannot have a
 matter-dominated phase if $n>1$ and if there is no interaction
 between the dark components in the universe~\cite{r33}.
 In~\cite{r33,r34}, a so-called ``new agegraphic dark energy''
 model was proposed to remove the inconsistency by replacing
 the time scale $T$ in Eq.~(\ref{eq5}) with the conformal
 time $\eta$. Of course, there exist other ways out of the
 difficulty in the original agegraphic dark energy model (see
 e.g. Sec.~2 of~\cite{r33}). Therefore, it is still worthwhile
 to study the original version of the agegraphic dark energy
 model. In this note, we will see that the interaction between
 the original agegraphic dark energy and pressureless (dark)
 matter can significantly change the cosmological evolution.
 Thus, the inconsistency in the original version without
 interaction can also be removed in the interacting agegraphic
 dark energy model.


\section{Interacting agegraphic dark energy}\label{sec2}

In this note, we extend the original agegraphic dark energy
 model by including the interaction between agegraphic dark
 energy and pressureless (dark) matter. Given the unknown
 nature of both dark energy and dark matter, it seems very
 peculiar that these two major components in the universe are
 entirely independent~\cite{r27,r28}. In fact, the models with
 interaction between dark energy and dark matter have been
 studied extensively in the literature. For a complete list of
 references concerning the interacting dark energy models, we
 refer to e.g.~\cite{r19,r20,r29} and references therein.

We assume that the agegraphic dark energy and pressureless (dark)
 matter exchange energy through an interaction term $Q$, namely
 \bea
 &&\dot{\rho}_q+3H\left(\rho_q+p_q\right)=-Q,\label{eq10}\\
 &&\dot{\rho}_m+3H\rho_m=Q,\label{eq11}
 \eea
 which preserves the total energy conservation equation
 $\dot{\rho}_{tot}+3H\left(\rho_{tot}+p_{tot}\right)=0$.
 In this work, we consider the three most familiar forms of
 interaction~\cite{r19,r20,r27,r28,r29}, namely
 \be{eq12}
 Q=3\alpha H\rho_q,~~~3\beta H\rho_m,~~~3\gamma H\rho_{tot},
 \ee
 where $\alpha$, $\beta$ and $\gamma$ are dimensionless constants.
 In fact, the interaction forms in Eq.~(\ref{eq12}) are given
 by hand. Although agegraphic dark energy is the quantum
 fluctuation of spacetime, it might decay into matter, similar
 to the $\Lambda(t)$CDM model in which the vacuum fluctuations
 can decay into matter. This effect could be described by the
 interaction term $Q$ phenomenologically. From Eq.~(\ref{eq7}),
 we get
 \be{eq13}
 \Omega_q^\prime=\Omega_q\left(-2\frac{\dot{H}}{H^2}
 -\frac{2}{n}\sqrt{\Omega_q}\right).
 \ee
 Differentiating Eq.~(\ref{eq6}) and using Eqs.~(\ref{eq11}),
 (\ref{eq5}) and~(\ref{eq7}), it is easy to find that
 \be{eq14}
 -\frac{\dot{H}}{H^2}=\frac{3}{2}\left(1-\Omega_q\right)
 +\frac{\Omega_q^{3/2}}{n}-\frac{Q}{6m_p^2 H^3}.
 \ee
 Therefore, we obtain the equation of motion for $\Omega_q$,
 \be{eq15}
 \Omega_q^\prime=\Omega_q\left[\left(1-\Omega_q\right)
 \left(3-\frac{2}{n}\sqrt{\Omega_q}\right)
 -\frac{Q}{3m_p^2 H^3}\right],
 \ee
 where
 \be{eq16}
 \frac{Q}{3m_p^2 H^3}=\left\{
 \begin{array}{ll}
 3\alpha\Omega_q & {\rm ~for~~} Q=3\alpha H\rho_q \\
 3\beta\left(1-\Omega_q\right) & {\rm ~for~~} Q=3\beta H\rho_m \\
 3\gamma & {\rm ~for~~} Q=3\gamma H\rho_{tot}
 \end{array}
 \right..
 \ee
 If $Q=0$, Eq.~(\ref{eq15}) reduces to Eq.~(\ref{eq8}). From
 Eqs.~(\ref{eq10}), (\ref{eq5}) and~(\ref{eq7}), we get the
 EoS of the agegraphic dark energy, namely
 \be{eq17}
 w_q=-1+\frac{2}{3n}\sqrt{\Omega_q}-\frac{Q}{3H\rho_q},
 \ee
 where
 \be{eq18}
 \frac{Q}{3H\rho_q}=\left\{
 \begin{array}{ll}
 \alpha & {\rm ~for~~} Q=3\alpha H\rho_q \\ \vspace{0.75mm}
 \beta\left(\Omega_q^{-1}-1\right) & {\rm ~for~~} Q=3\beta H\rho_m\\
 \gamma\,\Omega_q^{-1} & {\rm ~for~~} Q=3\gamma H\rho_{tot}
 \end{array}
 \right..
 \ee
 Again, if $Q=0$, Eq.~(\ref{eq17}) reduces to Eq.~(\ref{eq9}).
 Using Eq.~(\ref{eq14}), the deceleration parameter is given by
 \be{eq19}
 q\equiv -\frac{\ddot{a}a}{\dot{a}^2}=-1-\frac{\dot{H}}{H^2}
 =\frac{1}{2}-\frac{3}{2}\Omega_q
 +\frac{\Omega_q^{3/2}}{n}-\frac{Q}{6m_p^2 H^3},
 \ee
 where the last term can be found from Eq.~(\ref{eq16}). The
 total EoS $w_{tot}\equiv p_{tot}/\rho_{tot}=\Omega_q w_q$,
 where $w_q$ is given in Eq.~(\ref{eq17}). On the other hand,
 from the Friedmann equation and the Raychaudhuri equation,
 we have $w_{tot}=-1-\frac{2}{3}\frac{\dot{H}}{H^2}=-1/3+2q/3$.
 As mentioned above, in the case of $Q=0$ (i.e.~without
 interaction), $n>1$ is necessary to drive the (present)
 accelerated expansion of our universe. In the case of $Q\not=0$,
 the situation is changed. For example, if $Q=3\alpha H\rho_q$,
 to drive the (present) accelerated expansion of our universe,
 we should have $w_{tot}=\Omega_q w_q<-1/3$, which means that
 $n>2\Omega_q^{3/2}[3(1+\alpha)\Omega_q-1]^{-1}$. It is easy to
 see that the minimum of the right hand side of this inequality
 is $(1+\alpha)^{-3/2}$ at $\Omega_q=(1+\alpha)^{-1}$, if
 $\Omega_q>[3(1+\alpha)]^{-1}$~(nb. $\Omega_q\simeq 0.7$
 today). For $\alpha>0$, this minimum $(1+\alpha)^{-3/2}$ is
 smaller than $1$. For instance, if the present
 $\Omega_{q0}=0.7$, to drive the accelerated expansion of our
 universe today, $n>0.89414$ is enough for $\alpha=0.1$. In
 other words, $n$ can be smaller than $1$ to drive the
 accelerated expansion of our universe in the case of
 $Q\not=0$. We will see this point explicitly in the
 following~(e.g. Figs.~\ref{fig4} and~\ref{fig5}).

To get illustrations for the behaviors of $\Omega_q$, $w_q$,
 $q$ and $w_{tot}$, we show some numerical plots by using
 Eqs.~(\ref{eq15})---(\ref{eq19}) and $w_{tot}=\Omega_q w_q$.
 However, to be brief, we do not present plots for all forms
 of interaction $Q$. In the following, we mainly focus on
 the case of $Q=3\alpha H\rho_q$ as an example. Note that
 in the numerical integration of Eq.~(\ref{eq15}) we use
 the initial condition $\Omega_{q0}=0.7$ for demonstration.

\begin{center}
 \begin{figure}[htbp]
 \centering
 \includegraphics[width=0.75\textwidth]{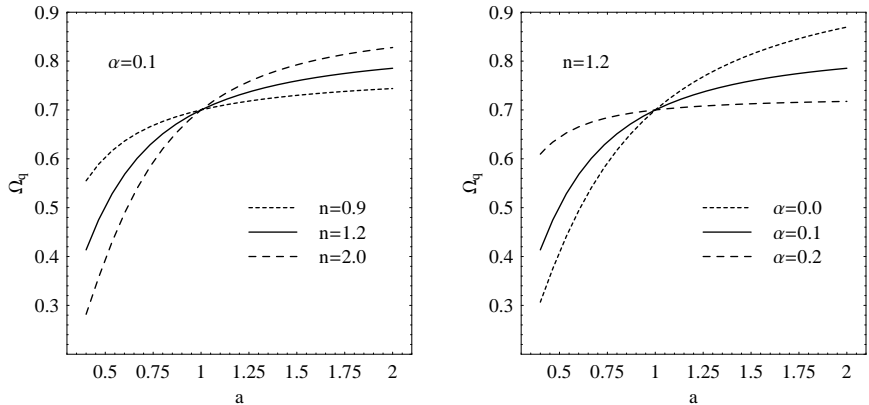}
 \caption{\label{fig1} Evolution of $\Omega_q$ for various
 model parameters $n$ and $\alpha$ in the case of
 $Q=3\alpha H\rho_q$.}
 \end{figure}
 \end{center}

In Fig.\ref{fig1}, we show the evolution of $\Omega_q$ for
 different model parameters $n$ and $\alpha$ in the case of
 $Q=3\alpha H\rho_q$. It is easy to see that for the fixed
 $\alpha$ which describes the interaction between the
 agegraphic dark energy and the pressureless (dark) matter,
 the agegraphic dark energy starts to be effective earlier
 and $\Omega_q$ tends to a lower value at the late time when
 $n$ is smaller. On the other hand, for fixed $n$, the
 agegraphic dark energy starts to be effective earlier and
 $\Omega_q$ tends to a lower value at the late time when
 $\alpha$ is larger. Interestingly enough, these behaviors are
 exactly opposite to the ones of the interacting holographic
 dark energy model~\cite{r21}. As mentioned above, this is
 mainly due to the opposite sign before $\sqrt{\Omega_{de}}$
 in the equation of motion for $\Omega_{de}$ (the subscript
 $de=q$ and $\Lambda$ for the agegraphic dark energy and
 holographic dark energy respectively).


 \begin{center}
 \begin{figure}[htbp]
 \centering
 \includegraphics[width=0.75\textwidth]{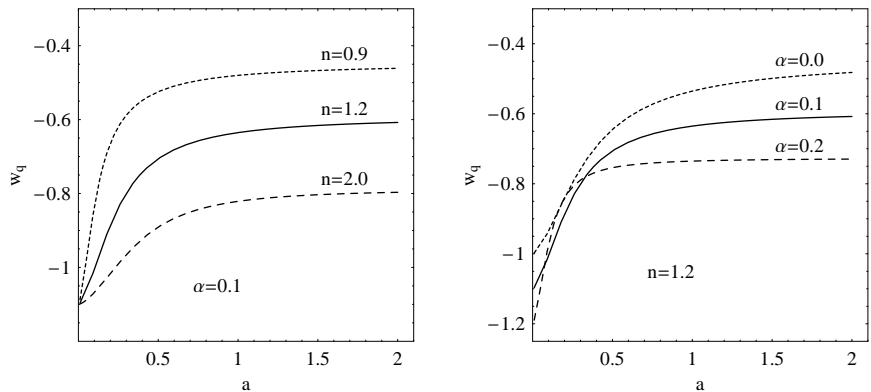}
 \caption{\label{fig2} Evolution of $w_q$ for various
 model parameters $n$ and $\alpha$ in the case of
 $Q=3\alpha H\rho_q$.}
 \end{figure}
 \end{center}


\vspace{-12mm} 

\begin{center}
 \begin{figure}[htbp]
 \centering
 \includegraphics[width=0.75\textwidth]{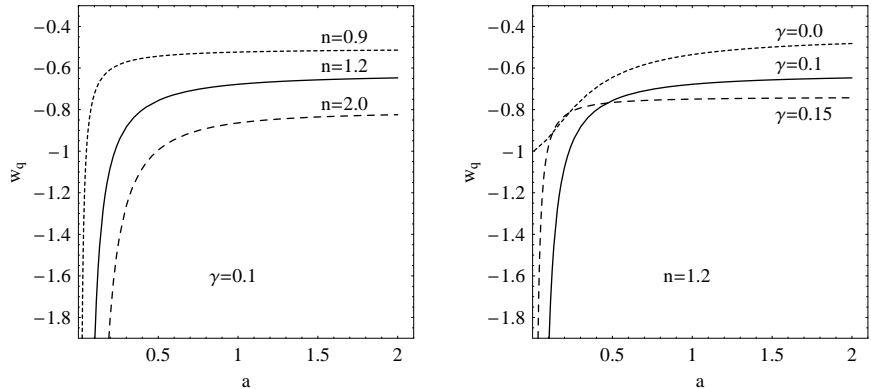}
 \caption{\label{fig3} Evolution of $w_q$ for various
 model parameters $n$ and $\gamma$ in the case of
 $Q=3\gamma H\rho_{tot}$.}
 \end{figure}
 \end{center}

\begin{center}
 \begin{figure}[tbp]
 \centering
 \includegraphics[width=0.75\textwidth]{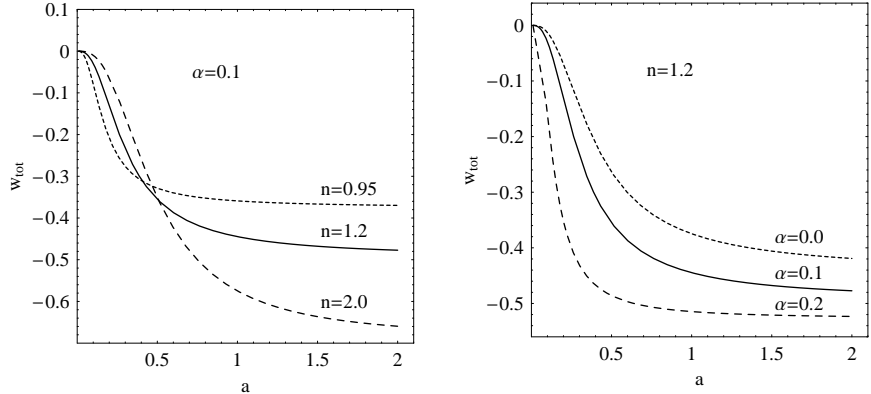}
 \caption{\label{fig4} Evolution of $w_{tot}$ for various
 model parameters $n$ and $\alpha$ in the case of
 $Q=3\alpha H\rho_q$.}
 \end{figure}
 \end{center}

\begin{center}
 \begin{figure}[tbp]
 \centering
 \includegraphics[width=0.75\textwidth]{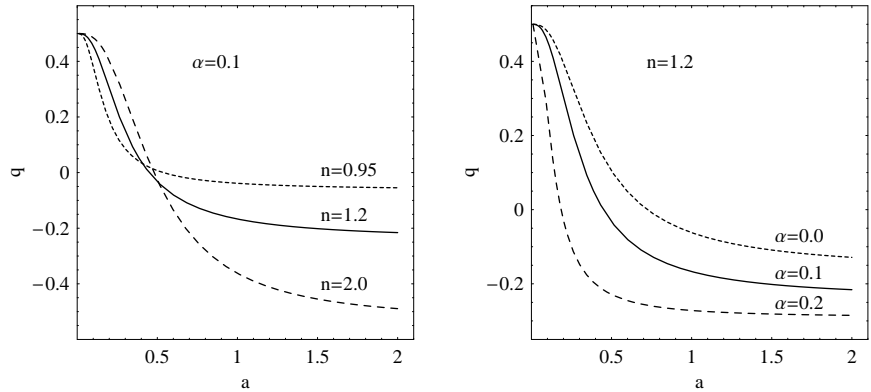}
 \caption{\label{fig5} Evolution of the deceleration parameter
 $q$ for various model parameters $n$ and $\alpha$ in the
 case of $Q=3\alpha H\rho_q$.}
 \end{figure}
 \end{center}

In Fig.~\ref{fig2}, we show the evolution of $w_q$ for different
 model parameters $n$ and $\alpha$ in the case of
 $Q=3\alpha H\rho_q$. Obviously, the EoS of the agegraphic dark
 energy $w_q$ can cross the phantom divide $w_{de}=-1$. In the
 case of $Q=0$ (i.e.~without interaction), as mentioned above,
 $w_q$ is always larger than $-1$ and cannot cross the phantom
 divide. By the help of interaction $Q\not=0$ between the
 agegraphic dark energy and the pressureless (dark) matter, the
 situation is changed. From Eq.~(\ref{eq17}) with the first line
 of Eq.~(\ref{eq18}), it is easy to understand that $w_q$
 converges to the value $-1-\alpha$ at the early time in the case
 of $Q=3\alpha H\rho_q$. Of course, the most interesting
 observation from Fig.~\ref{fig2} is that $w_q$ crosses the
 phantom divide from $w_q<-1$ to $w_q>-1$. This makes it
 distinguishable from many other dark energy models whose $w_{de}$
 can cross the phantom divide. It can be categorized into the
 so-called Quintom~B type model, in the terminology
 of~\cite{r22,r23}. To make this point more robust, we also
 plot the evolution of $w_q$ for various model parameters $n$
 and $\gamma$ in the case of $Q=3\gamma H\rho_{tot}$. The
 results are presented in Fig.~\ref{fig3}. Clearly, the observation
 that $w_q$ crosses the phantom divide from below to above still
 holds. By the way, the $w_q$ tends to $-\infty$ at early time
 for $\gamma\not=0$; this is due to the last term in
 Eq.~(\ref{eq17}) with the last line of Eq.~(\ref{eq18}) in the
 case of $Q=3\gamma H\rho_{tot}$. It is anticipated that the
 behavior of $w_q$ in the case of $Q=3\beta H\rho_m$ is similar
 to the case of $Q=3\gamma H\rho_{tot}$, since the last terms in
 the versions of Eq.~(\ref{eq17}) for these two cases are
 similar [cf. Eq.~(\ref{eq18})]. It is worth noting that these
 results are for the cases of positive $\alpha$, $\beta$ and
 $\gamma$. In the cases of negative $\alpha$, $\beta$ and
 $\gamma$, from Eq.~(\ref{eq17}) with Eq.~(\ref{eq18}), one can
 see that $w_q$ is always larger than $-1$ and cannot cross the
 phantom divide. Obviously, the cases of positive $\alpha$,
 $\beta$ and $\gamma$ are more interesting since the $w_q$
 can cross the phantom divide from $w_q<-1$ to $w_q>-1$.


\vspace{-5mm} 

One of the benefits of $w_q>-1$ at late time in the
 interacting agegraphic dark energy model is that the
 universe can avoid the big rip singularity~\cite{r24,r25}.
 Of course, the direct condition for the avoidance of
 big rip should be $w_{tot}>-1$ instead. Since
 $w_{tot}=\Omega_q w_q$ and $0\leq\Omega_q\leq 1$, the
 condition $w_{tot}>-1$ is automatically satisfied when
 $w_q>-1$. This can be seen clearly from the plot of the
 evolution of $w_{tot}$ for various model parameters $n$
 and $\alpha$ in the case of $Q=3\alpha H\rho_q$ for example,
 which is shown in Fig.~\ref{fig4}.

The other thing one can see from Fig.~\ref{fig4} is that
 $w_{tot}>-1/3$ at the early time and $w_{tot}<-1/3$ at the
 late time. This implies that the universe undergoes
 decelerated expansion at the early time and later starts
 accelerated expansion. To see this point clearly, we show the
 deceleration parameter $q$ in Fig.~\ref{fig5}. Obviously,
 $q$ crosses the boundary $q=0$ from $q>0$ to $q<0$. Some
 remarks on Figs.~\ref{fig4} and~\ref{fig5} are in order.
 First, the similarity between these two figures is due to
 the relation $w_{tot}=-1/3+2q/3$ mentioned above. Second,
 at the early time $w_{tot}$ and $q$ converge to $0$ and $1/2$,
 respectively; this is because $\Omega_q$ can be neglected at
 the early time, whereas the universe is dominated by
 the pressureless (dark) matter. This is in the case of
 $Q=3\alpha H\rho_q$. In the cases of $Q=3\beta H\rho_m$
 and $Q=3\gamma H\rho_{tot}$, however, from Eqs.~(\ref{eq19}),
 (\ref{eq17}) and $w_{tot}=\Omega_q w_q$, one can see that
 at the early time $w_{tot}$ and $q$ converge to other
 constants rather than $0$ and $1/2$. For instance, in the case
 of $Q=3\beta H\rho_m$, at the early time $w_{tot}\to -\beta$
 and $q\to 1/2-3\beta/2$. In the case of
 $Q=3\gamma H\rho_{tot}$, at early time $w_{tot}\to -\gamma$
 and $q\to 1/2-3\gamma/2$. Third, for fixed $n$, the universe
 starts accelerated expansion earlier when $\alpha$ is larger
 (see the right panels of Figs.~\ref{fig4} and~\ref{fig5}).
 Fourth, the universe will undergo accelerated expansion at
 the late time forever and cannot come back to decelerated
 expansion, as shown in Figs.~\ref{fig4} and~\ref{fig5}. After
 all, in the case of $Q=3\alpha H\rho_q$, we notice that for
 $\alpha=0.1$, the universe can undergo accelerated expansion
 for $n=0.95<1$. One can see this point from Figs.~\ref{fig4}
 and~\ref{fig5}. As mentioned above, this is impossible in the
 case of $Q=0$ (i.e.~without interaction). The interaction
 $Q\not=0$ changes the situation.

Before the end of this work, we would like to mention another
 interesting feature of the interacting agegraphic dark energy
 model. The equation of motion for $\Omega_q$,
 Eq.~(\ref{eq15}), can be viewed as an one-dimensional
 dynamical system~\cite{r26}. The critical points of this
 autonomous equation are determined by $\Omega_q^\prime=0$.
 They are $\Omega_q=0$ and the solutions of the equation
 \be{eq20}
 \left(1-\Omega_q\right)
 \left(3-\frac{2}{n}\sqrt{\Omega_q}\right)
 =\frac{Q}{3m_p^2 H^3},
 \ee
 where the right hand side is given in Eq.~(\ref{eq16}).
 In the case of $Q=0$ (i.e.~without interaction), the physical
 solution of Eq.~(\ref{eq20}) is only $\Omega_q=1$, whereas the
 other solution $\sqrt{\Omega_q}=3n/2>1$ is unphysical because
 $n>1$ is required by the accelerated expansion of our universe,
 as mentioned above. Thus, there is no scaling solution in the
 case without interaction. Again, this situation is changed
 in the cases of $Q\not=0$. For instance, in the case of
 $Q=3\beta H\rho_m$, the critical points are $\Omega_q=0$,
 $1$ and $9n^2 (1-\beta)^2/4$. Note that in the case of
 $Q\not=0$, $n>1$ is not necessary to drive the accelerated
 expansion of our universe. We can choose appropriate model
 parameters $n$ and $\beta$ to ensure
 $0<\Omega_q=9n^2 (1-\beta)^2/4=const.<1$. Thus, there is a
 scaling solution. In the cases of $Q=3\alpha H\rho_q$ and
 $Q=3\gamma H\rho_{tot}$, the situation is similar. The
 solutions are fairly complicated and we do not present them
 here, since Eq.~(\ref{eq20}) is a cubic equation of
 $\sqrt{\Omega_q}$ in these two cases. In fact, the flat tails
 of some curves in Figs.~\ref{fig1}---\ref{fig5} perhaps hint
 the scaling solutions in the late time. The scaling solutions
 in the cases of $Q\not=0$ can help to alleviate the
 coincidence problem. As is well known, for a dynamical
 system~\cite{r26}, the universe will enter the attractors in
 the late time, regardless of the initial conditions. If the
 attractors are scaling solutions, both $\Omega_q$ and
 $\Omega_m=1-\Omega_q$ are fixed values over there. If $n$ and
 $\alpha$, $\beta$, $\gamma$ are of order unity, it is not
 surprising that $\Omega_q$ and $\Omega_m=1-\Omega_q$ are
 comparable at the late time. In fact, this is just the
 essential point of the literature (see
 e.g.~\cite{r19,r20,r27,r28,r29}) to {\em alleviate} (rather
 than {\em solve}) the coincidence problem using the method of
 dynamical system~\cite{r26}.


\vspace{-18mm} 


\section{Concluding remarks}\label{sec3}

In summary, we have extended the agegraphic dark energy
 model by including the interaction between the agegraphic
 dark energy and the pressureless (dark) matter. The original
 agegraphic dark energy model was proposed in~\cite{r17} based
 on the K\'{a}rolyh\'{a}zy uncertainty relation, which arises
 from quantum mechanics together with general relativity.
 In the interacting agegraphic dark energy model, there are
 many interesting features different from the original
 agegraphic dark energy model and holographic dark energy
 model. In the cases with interaction $Q\not=0$, the parameter
 $n>1$ is no longer necessary to drive the accelerated
 expansion of our universe; the EoS of agegraphic dark energy
 can cross the phantom divide, whereas the big rip can be
 avoided; the universe undergoes decelerated expansion at early
 time and then starts accelerated expansion later; there are
 scaling solutions which can help to alleviate the coincidence
 problem. In particular, the difficulty in the original version
 of the agegraphic dark energy model~\cite{r17} can be avoided
 here, thanks to the interaction between the dark components.

It is of interest to discuss the similarity and difference
 between agegraphic dark energy and holographic dark energy.
 It is shown that the agegraphic dark energy naturally obeys the
 holographic black hole entropy bound~\cite{r12,r17}, just like
 holographic dark energy. By choosing the age of the universe
 rather than the future event horizon as the length measure, the
 drawback concerning causality in the holographic dark energy model
 does not exist in the agegraphic dark energy model~\cite{r17}.
 It is worth noting that the agegraphic energy density
 Eq.~(\ref{eq3}) is similar to the one of holographic dark
 energy~\cite{r14,r13,r16,r15}, i.e.,
 $\rho_\Lambda\sim l_p^{-2}l^{-2}$. The similarity between
 $\rho_q$ and $\rho_\Lambda$ might reveal some universal features
 of quantum gravity, although they arise in different ways. In
 addition, the sign before the term $\sqrt{\Omega_{de}}$~(the
 subscript $de=q$ and $\Lambda$ for agegraphic dark energy and
 holographic dark energy, respectively) is opposite in the equation
 of motion for $\Omega_q$, the EoS of the agegraphic dark energy
 $w_q$, the total EoS $w_{tot}$ and the deceleration parameter $q$.
 This difference brings about some interesting features to the
 agegraphic dark energy different from the ones of holographic
 dark energy. In some sense, the relation between agegraphic
 dark energy and holographic dark energy is similar to the
 one between phantom and quintessence.

Finally, some remarks are in order. First, we admit that a
 sufficiently strong interaction might be required to relax the
 condition $n>1$ for an accelerated expansion and to allow that
 $w_q$ crosses the phantom divide. However, a strong
 interaction might encounter fairly tight constraints from
 local gravity tests. Second, after the appearance of our
 relevant works on the (new) agegraphic dark energy, it was
 found that the original agegraphic dark energy model proposed
 in~\cite{r17} is difficult to reconcile with the big bang
 nucleosynthesis (BBN) constraint~\cite{r35}. On the other
 hand, as shown in~\cite{r36}, the situation is better in the
 new agegraphic dark energy model~\cite{r33,r34}. In addition,
 the (new) agegraphic dark energy model faces the problem of
 instabilities~\cite{r37}, while the holographic dark energy
 model also faces the same problem~\cite{r38}. Third, the
 quintessence reconstructions of the (new) agegraphic dark
 energy have been studied in~\cite{r39}. The statefinder
 diagnostic and $w-w^\prime$ analysis for the agegraphic dark
 energy models were performed in~\cite{r40}. In addition, the
 (new) agegraphic dark energy was extended with the generalized
 uncertainty principle in~\cite{r41}. Furthermore, it was
 argued that the holographic dark energy models might share
 the same origin with the (new) agegraphic dark energy
 models~\cite{r42}. So, we consider the (new) agegraphic dark
 energy model to deserve further investigation in future work.


\section*{ACKNOWLEDGMENTS}
We are grateful to Prof.~Miao~Li, Prof.~Shuang~Nan~Zhang and
 Prof.~Yi~Ling for helpful discussions. We also thank Minzi~Feng,
 as well as Hui~Li, Yi~Zhang, Xing~Wu, Xin~Zhang, Jingfei~Zhang,
 Jian-Pin~Wu, Jian~Wang and Bin~Hu, for kind help and discussions.
 This work was supported in part by a grant from China Postdoctoral
 Science Foundation, a grant from Chinese Academy of
 Sciences~(No.~KJCX3-SYW-N2), and by NSFC under grants No.~10325525,
 No.~10525060 and No.~90403029.

\renewcommand{\baselinestretch}{1.1}


\end{document}